\documentclass[aps,prb,reprint,showpacs,citeautoscript,
               longbibliography]{revtex4-1}

\usepackage{amsmath,amssymb,amsfonts,graphicx}

\newcommand*{\half}{\frac{1}{2}}

\newcommand*{\avg}[1]{\langle #1 \rangle}

\begin{document}

\title{Cavity Quantum Electrodynamics of a two-level atom
       with modulated fields}
\author{U. Pishipati}
\author{I. Almakremi}
\author{Amitabh Joshi}
\email{ajoshi@eiu.edu}
\affiliation{Department of Physics, Eastern Illinois University,
             Charleston, Illinois 61920}
\author{Juan D. Serna}
\email{serna@uamont.edu}
\affiliation{School of Mathematical and Natural Sciences,
             University of Arkansas at Monticello, Monticello, Arkansas 71656}

\date{July 28, 2011}

%% -----------------------------------------------------------------------------
%%                                 ABSTRACT
%% -----------------------------------------------------------------------------
\begin{abstract}
We studied the interaction of a two-level atom with a frequency modulated cavity
mode in an ideal optical cavity. The system, described by a Jaynes-Cumming
Hamiltonian, gave rise to a set of stiff nonlinear first order equations solved
numerically using implicit and semi-implicit numerical algorithms. We explored
the evolution of the atomic system using nonlinear dynamics tools, like time
series, phase plane, power spectral density, and Poincar\'{e} sections plots,
for monochromatic and bichromatic modulations of the cavity field. The system
showed quasiperiodic and possibly chaotic behavior when the selected
monochromatic frequencies, or ratio of the bichromatic frequencies were
irrational (incommensurate) numbers. In addition, when the modulated frequencies
were overtones of the Rabi frequency of the system, a single dominant frequency
emerged for the system.
\end{abstract}


\maketitle

%% -----------------------------------------------------------------------------
%%                               INTRODUCTION
%% -----------------------------------------------------------------------------
\section{\label{sec:intro}Introduction}

There has been a considerable amount of interest in the study of simple quantum
mechanical systems, such as the simple harmonic oscillator or the two-level atom
interacting with quasiperiodic electromagnetic fields. Initially, attention was
directed to understand quantum chaos
phenomenon~\cite{Casati,Bohigas:52,Casati:95}. Later on, it was observed that
the atomic interaction with quasiperiodic fields could result into some other
novel dynamical features like power broadening and resonance
shifts~\cite{Ho:17}, multiple resonance spectral structures~\cite{Agarwal:85},
population trapping~\cite{Raghavan:54}, sinusoidal and square wave oscillation
of the population~\cite{Noel:58,Zhang:68}, and new types of dynamical
localization~\cite{Noba:64}. It was also noted that the atomic response to
quasiperiodic fields is extremely complex, and the diagnostic techniques
employed for characterizing chaos could easily be utilized for studying the
complexities of such interactions.

A good number of studies are available in which investigations of atomic systems
interacting with a quasiperiodic amplitude modulated field have been carried
out, finding that the atomic density matrix elements have rapidly decaying
autocorrelation functions~\cite{Milonni}. In one of these
works~\cite{Smirnov:52}, the dynamics of a two-level atom interacting with a
quasiperiodic field was examined as a function of the field strength (which
could be considered as a controlling parameter). Also, the atomic dynamics was
characterized through the computation of Poincar\'{e} sections of the Bloch
vectors's motion, and the attractor's fractal nature~\cite{Camparo:43}. Another
interesting work involved the situation in which a single two-level atom was
interacting with a combination of a classical and a quantized fields in the
cavity, so that the cavity field acted as a quantum probe of the dressed states
defined by the classical laser field~\cite{Law:43}. Many interesting effects
such as two-photon gain, cavity perturbed resonance fluorescence spectra, and
the atomic squeezing in the cavity were examined. Chaotic Rabi oscillations
under quasiperiodic perturbation were also studied~\cite{Pomeau:56}, where a
system consisting of a two-level atom was coupled through a time-dependent,
quasiperiodic perturbation with two incommensurate frequencies. In this system,
chaos was predicted by observing a rapidly decreasing autocorrelation and by
looking at the continuous Fourier spectrum of the time-dependent physical
observable. More recently, the use of time-dependent dissipative environments to
control the quantum state of a two-level atom was
demonstrated~\cite{Linington:77}. For this case, the possibility to decouple
dynamically the atom from its environment leading to Markovian dynamics was
studied.

In this paper, we show an alternative approach to explore the response of an
atomic system to a quasiperiodic field through its interaction with a modulated
single mode field sustained in an ideal cavity. The frequency of the cavity
field is modulated while its amplitude is kept constant, making this work
different from the studies reported earlier. The interaction of a modulated
field with a two-level atom is analyzed by studying the dynamical evolution,
phase plot, Poincar\'{e} section, and power spectrum density of the system under
certain parametric conditions. We solve the set of nonlinear differential
equations that describe the evolution of the system using implicit and
semi-implicit algorithms specialized in dealing with stiff differential
equations. We observe that for specific frequency values, the system's evolution
is quasiperiodic, and the possibility of chaos is present. The cavity field is
first modulated monochromatically, and secondly bichromatically such that the
frequencies involved and their ratios are either irrational numbers or integral
multiples of the Rabi frequency of the system. The paper is organized as
follows. In section~\ref{sec:theModel}, we describe the physical model of a
two-level atom interacting with a modulated quantized cavity-field mode.
Section~\ref{sec:results} is devoted to results and discussions of the studied
atomic system. Finally, some concluding remarks are presented in
section~\ref{sec:summary}.

%% -----------------------------------------------------------------------------
%%                            THEORY AND METHODS
%% -----------------------------------------------------------------------------
\section{\label{sec:theModel}Model and basic equations}

We consider the interaction of a two-level atom at rest, and atomic transition
frequency $\omega_0$, with a quasiperiodic quantized cavity field $\omega_c(t)$.
The quasiperiodic (or incommensurate) field is obtained by modulating the
quantized cavity field frequency. The Hamiltonian of the system under the
rotating wave approximation is
\begin{equation}\label{Eq:Hamiltonian}
  H = \omega_{0}S_{z} + \omega_c(t)\,a^{\dagger}a
      + g(S_{+}a + a^{\dagger}S_{-}),
\end{equation}
where $\hbar=1$ for simplicity, $a$ and $a^{\dagger}$ are the annihilation and
creation operators for the cavity field, and $g$ is the atom field coupling
coefficient; $S_{z}$ and $S_{\pm}$ are the usual spin-$\half$ operators
satisfying the commutation relations
\begin{equation}\label{Eq:commutators}
  [S_z, S_{\pm}] = \pm S_{\pm}, \qquad [S_{+}, S_{-}] = 2 S_z.
\end{equation}
If the modulation of the cavity field is slow, that is, the variation of
$\omega_c(t)$ with respect to $\omega_0$ is small, then the operator
$a^{\dagger}a+S_z$ is a constant of motion, and the \textit{interaction}
picture Hamiltonian of the system takes the form
\begin{equation}\label{Eq:IHamiltonian}
  H_{I} = \delta(t) S_z + g(S_{+} a + a^{\dagger} S_{-}),
\end{equation}
where $\delta (t)= \omega_0 - \omega _c(t)$. In this picture, the operators
incorporate a dependency on time, and the equation of motion for an arbitrary
time-dependent operator $q$ is given by
\begin{equation}\label{Eq:Heisenberg}
  i \dot{q} = [q, H_{I}].
\end{equation}
Hence the Heisenberg's equations of motion for the atomic and field operators of
Hamiltonian~\eqref{Eq:IHamiltonian} are
\begin{equation}\label{Eq:system01}
\begin{split}
  i \dot{a}   &= gS_{-}, \\
  i \dot{S}_z &= g(S_{+} a - a^{\dagger} S_{-}), \\
  i \dot{S}_+ &= -\delta S_{+}+2ga^{\dagger}S_z, \\
  i \dot{S}_- &= \delta S_{-}-2gaS_z.
\end{split}
\end{equation}
Equation~\eqref{Eq:system01} gives rise to a \textit{nonlinear} third order
differential equation in $S_z$
\begin{equation}\label{Eq:dddSz}
\begin{split}
  \dddot{S}_z - \dfrac{\dot{\delta}}{\delta}\,\ddot{S}_z
  + [\delta^{2} +{}& 4g^{2}(N+1/2)]\,\dot{S}_z \\
  - {}& 4g^{2}(N+1/2)\,\dfrac{\dot{\delta}}{\delta}\,S_z = 0,
\end{split}
\end{equation}
where $N \equiv a^{\dagger}a + S_z$ is a constant of motion.

If we define the following new variables
\begin{equation}\label{Eq:system02}
\begin{split}
  y_1 &= \avg{S_z}, \\
  y_2 &= \dot{y}_1 = \avg{\dot{S}_z}, \\
  y_3 &= \dot{y}_2 = \avg{\ddot{S}_z},
\end{split}
\end{equation}
then it is possible to recast~\eqref{Eq:dddSz} into the equivalent system of
three, nonlinear first order differential equations
\begin{equation}\label{Eq:system03}
\begin{split}
\dot {y_1} = {}& y_2, \\
\dot {y_2} = {}& y_3, \\
\dot {y_3} = {}& \dfrac{\dot{\delta}}{\delta}\,y_3
              - [\delta^{2} + 4g^{2}(N+1/2)]\,y_2 \\
             {}& + 4g^{2}(N+1/2)\,\dfrac{\dot{\delta}}{\delta}\,y_1.
\end{split}
\end{equation}
Solving either~\eqref{Eq:system01} or~\eqref{Eq:system03} numerically, the
dynamical evolution of the system can easily be obtained for different
parametric conditions.

The system of nonlinear differential equations~\eqref{Eq:system03} includes some
terms ($\dot{\delta} / \delta$ and $\delta^2$) that lead to rapid oscillations
of the solution. This makes the set of equations \textit{stiff}, so very stable
numerical algorithms must be used to find its solutions. Because this system
was small in size (only three differential equations) and we looked for
solutions with moderate accuracies, we used \textit{implicit} Runge-Kutta, and
\textit{semi-implicit} Rosenbrock methods to solve the system~\cite{NR:C++}.

We estimated the \textit{power spectral density} (PSD) for each one of the
studied cases to identify periodicities, dominant frequencies, and their
correlation with the Rabi frequency $\Omega$ of the system. To obtain these
power densities, we used a simple \textit{periodogram} estimator by taking an
$n$-point sample of the function $\avg{S_z(t)}$ at equal intervals $\Delta$, and
using the fast Fourier transform to compute its \textit{forward} discrete
Fourier transform~\cite{FFTW}
\begin{equation}\label{Eq:FFT}
  Y_k = \sum_{j=0}^{n-1} \avg{S_z}_j\,e^{-2\pi j k \sqrt{-1}/ n} \qquad
  k = 0, 1, 2, \ldots\,.
\end{equation}
Then the power spectrum was defined at $n/2+1$ frequencies as
\begin{equation}\label{Eq:PSD}
\begin{split}
  P(f_k) ={}& \frac{1}{n^2}\left[ |Y_k|^2 + |Y_{n-k}|^2 \right] \\
       k ={}& 1, 2, \ldots, \left(\frac{n}{2}-1\right),
\end{split}
\end{equation}
where $f_k$ took values only for zero and positive frequencies
\begin{equation}\label{Eq:PSDfrequency}
  f_k = \frac{\omega_k}{2\pi} = \frac{k}{n \Delta} \qquad \quad
  k = 0, 1, \ldots, \frac{n}{2}.
\end{equation}

Finally, to facilitate the comparison and analysis of the dynamics of both the
monochromatic and bichromatic modulated systems, we calculated the Poincar\'{e}
sections (PS) by taking stroboscopic views of the phase plane ($\avg{S_z}$,
$\avg{\dot{S}_z}$) for just the values of
\begin{equation}\label{Eq:Poincare}
 g t = \frac{2\pi\,n}{\sqrt{\delta_0^2 + 4g^2(N+1/2)}} \qquad
 n = 0, 1, 2,\ldots \,,
\end{equation}
corresponding to periods equal to the Rabi period of the systems. In this
equation, $\delta_0 \equiv \delta(t=0)$.

%% -----------------------------------------------------------------------------
%%                          RESULTS AND DISCUSSION
%% -----------------------------------------------------------------------------
\section{\label{sec:results}Results and Discussion}

We solved the set of equations~\eqref{Eq:system03} numerically, using the
implicit Runge-Kutta and semi-implicit Rosenbrock methods~\cite{NR:C++}. The
algorithms are adaptive (variable stepsize $h$), and the absolute and relative
tolerance errors (\texttt{atol} and \texttt{rtol}) can be varied in accordance
to the pursued accuracy, and the system of differential equations. Although the
initial guess for the stepsize and absolute tolerance error we used in some of
the calculations were relatively small ($h \sim 10^{-7}$ and
$\mbox{\texttt{atol}} \sim 10^{-9}$), the algorithms showed stability and
reliability. The results obtained for each set of parameters used were
consistent and within the expected error. However, it is necessary to mention
that, for monochromatic modulations, the algorithm displayed some instability
and erratic behavior when the modulated frequencies were smaller than the Rabi
frequency of the system. This behavior may have to do with the very fast
oscillation of the solutions when the frequency differences are nearly zero, and
they are also present as denominators in the system of differential equations.

\begin{figure}[b]
\centering
\includegraphics[scale=0.16]{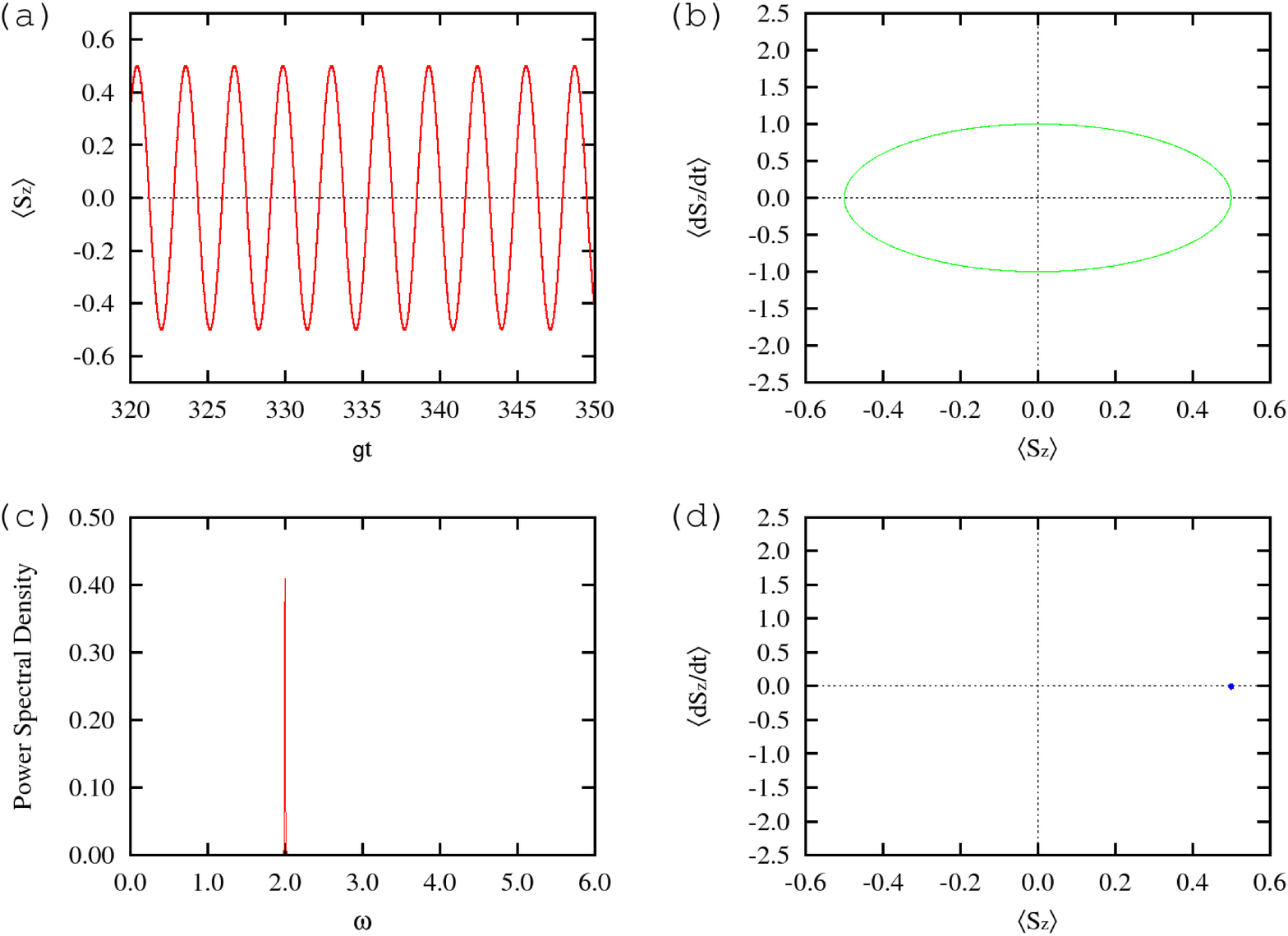}
\caption{\label{Fig01}Atom-cavity interaction for a fixed quantized cavity
field. The parameters of the system are: $N=3.50$, $g=1.00$, $\delta_0=0$, and
$\avg{S_{z}(0)}=+1/2$. The plots are: (a) time series, (b) phase plane, (c)
power spectral density, and (d) Poincar\'{e} section.}
\end{figure}

We first consider the case of no modulation of the quantized cavity field. In
Fig.~\ref{Fig01}(a), we plot $\avg{S_{z}(t)}$ as a function of the scaled time
$gt$, this is the \textit{time series} for $\avg{S_{z}(t)}$. The parameters
selected were $N=3.50$, $g=1.00$, $\avg{S_{z}(0)}=+1/2$ (the atom is initially
in its excited state), and $\delta_0=0$ (there is no modulation for the cavity
field). The time series is a sinusoidal curve representing how the population of
the two-level atom evolves when excited with the Fock state cavity field. The
mathematical behavior of $\avg{S_{z}(t)}$ was quite straightforward in this
case. Figure~\ref{Fig01}(b) shows the \textit{phase plane} plot of
$\avg{\dot{S}_z(t)}$ against $\avg{S_z(t)}$. The phase trajectory is a
\textit{closed} curve with a single period of evolution. The PSD corresponding
to the time series is shown in Fig.~\ref{Fig01}(c). We observe there is just one
single frequency in the time series (Fig.~\ref{Fig01}(a)) and this concur with
the result observed in Fig.~\ref{Fig01}(c), where we can see only one peak in
the Fourier spectrum. This frequency matches the Rabi frequency of the system,
which is equal to $4.0$ in $g^{-1}$ units. As expected, there was only a single
point in the PS plot (Fig.~\ref{Fig01}(d)). Here, the PS represents when the
phase trajectory crosses the time plane at every Rabi period. When we changed
the parameter $\delta_0$ to a value different from zero, all the features
displayed in Fig.~\ref{Fig01}(a-d) remained unchanged, except that the Rabi
frequency now turns into a more general expression given by $[\delta_0^{2} +
4g^{2}(N+1/2)]^{1/2}$. In addition, when we changed the value of $N$, the
generalized Rabi frequency changed. However, the features represented in
Fig.~\ref{Fig01} still remained unchanged (single frequency in the time series,
closed curve in the phase plane, and single point in the PS).

\begin{figure}[b]
\centering
\includegraphics[scale=0.16]{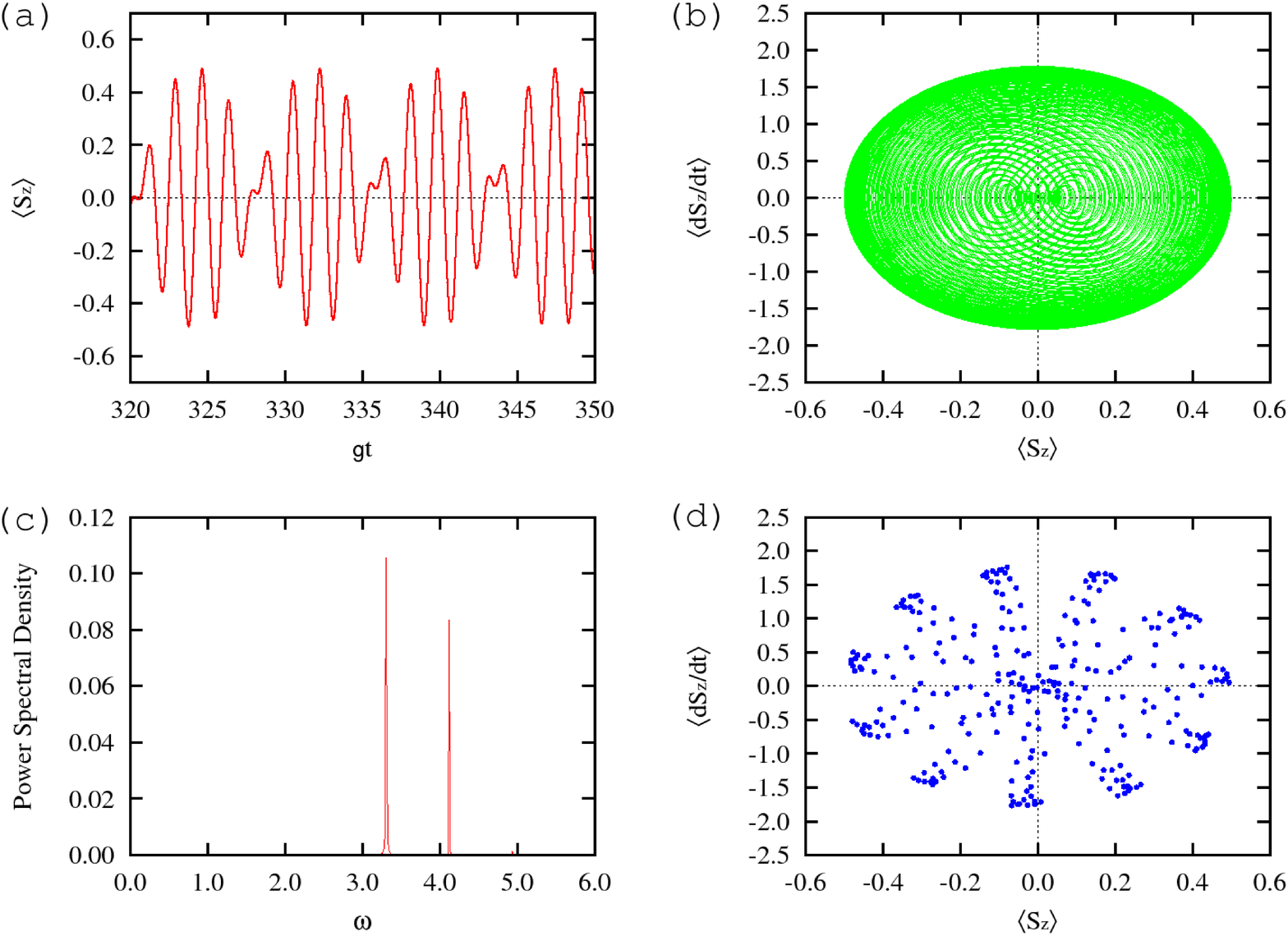}
\caption{\label{Fig02}Monochromatic modulation of the cavity field. The time
series shows a modulation pattern. Two dominant frequencies are present in the
power spectrum plot. The parameters of the system are: $N=2.50$, $g=1.00$,
$\delta_0=1.00$, $\omega=\sqrt{17}$, and $\avg{S_{z}(0)}=+1/2$. The plots are:
(a) time series, (b) phase plane, (c) power spectral density, and (d)
Poincar\'{e} section.}
\end{figure}

For the next set of parameters, we introduced mono\-chromatic modulation of the
cavity field in the form $$\delta=\delta_0\cos(\omega t).$$ We chose $N=2.50$,
$g=1.00$, $\delta_0=1.00$, and $\omega=\sqrt{17}$. The corresponding Rabi
frequency was $\Omega = 3.60$ (see Fig.~\ref{Fig02}(a-d)). In this case, we
modulated the cavity field with a single frequency $\omega=\sqrt{17}=4.12$,
which is an \textit{irrational} number. This fact gave rise to an interesting
behavior of the dynamical evolution of the system. Figure~\ref{Fig02}(a) shows
the time series for $\avg{S_{z}(t)}$ which clearly exhibits a modulation
pattern. The phase plot area is almost filled by the phase trajectory from zero
to the maxima of the two variables. The PSD showed two peaks, one located around
$\Omega$ (the Rabi frequency of the system) and the other at $\omega$ (the
modulated frequency of the cavity). The PS is shown in Fig.~\ref{Fig02}(d).
Although there is a lack of synchronization between the two frequencies in the
system, a regular star-like pattern formed, showing that some regions of the
phase plane (states of the system) were visited more often than others,
suggesting the existence of possible \textit{attractors}.

\begin{figure}[b]
\centering
\includegraphics[scale=0.16]{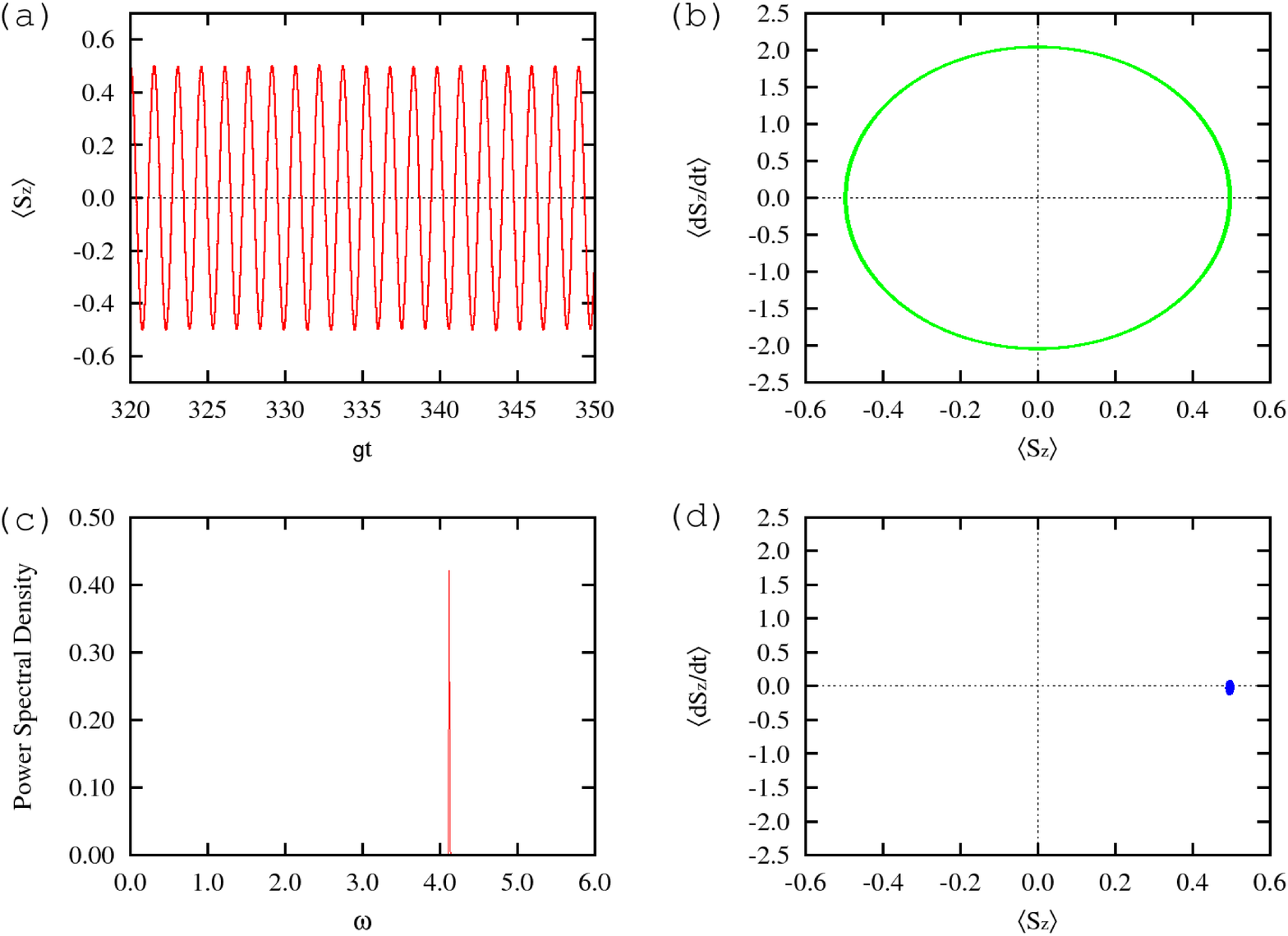}
\caption{\label{Fig03}The monochromatic modulation frequency of the cavity
equals the Rabi frequency of the system. The parameters of the system are:
$N=3.50$, $g=1.00$, $\delta_0=1.00$, $\omega=\Omega=\sqrt{17}$, and
$\avg{S_{z}(0)}=+1/2$. The plots are: (a) time series, (b) phase plane, (c)
power spectral density, and (d) Poincar\'{e} section. A similar situation is
presented when $\omega = m\,\Omega$, ($m=1,2,3,\ldots$).}
\end{figure}

Interestingly, when $\omega=\Omega=\sqrt{17}$, the situation changed drastically
as shown in Fig.~\ref{Fig03}. We achieved this condition for $N = 3.50$. The
time series now shows a single frequency (see Fig.~\ref{Fig03}(a)) which is in
accordance with the PSD of Fig.~\ref{Fig03}(c). The phase plot is again a closed
curve as depicted in Fig.~\ref{Fig03}(b), whereas the PS shows a number of dots
making up a wide single point in Fig.~\ref{Fig03}(d). These points are supposed
to be at the same location; however, their separations are caused by the
propagation of the computational error. In this case, when the condition
$\omega=\Omega$ is satisfied, the cavity modulation of the system synchronizes
with the Rabi frequency of the system, and there is a single frequency or single
period of oscillation in the dynamical evolution. We also observed another
intriguing situation when we chose the modulation frequency to be an overtone
(integral multiple) of the Rabi frequency, i.e., $\omega = m\,\Omega$,
($m=1,2,3,\ldots$). For this situation, we observed the same synchronized
behavior shown in Fig.~\ref{Fig03}. This phenomenon happens no matter the value
of $\Omega$, whether it is an integer, rational, or irrational number. When
$m\,\Omega < \omega < (m+1)\,\Omega$, this synchronization is lost.

\begin{figure}[t]
\centering
\includegraphics[scale=0.16]{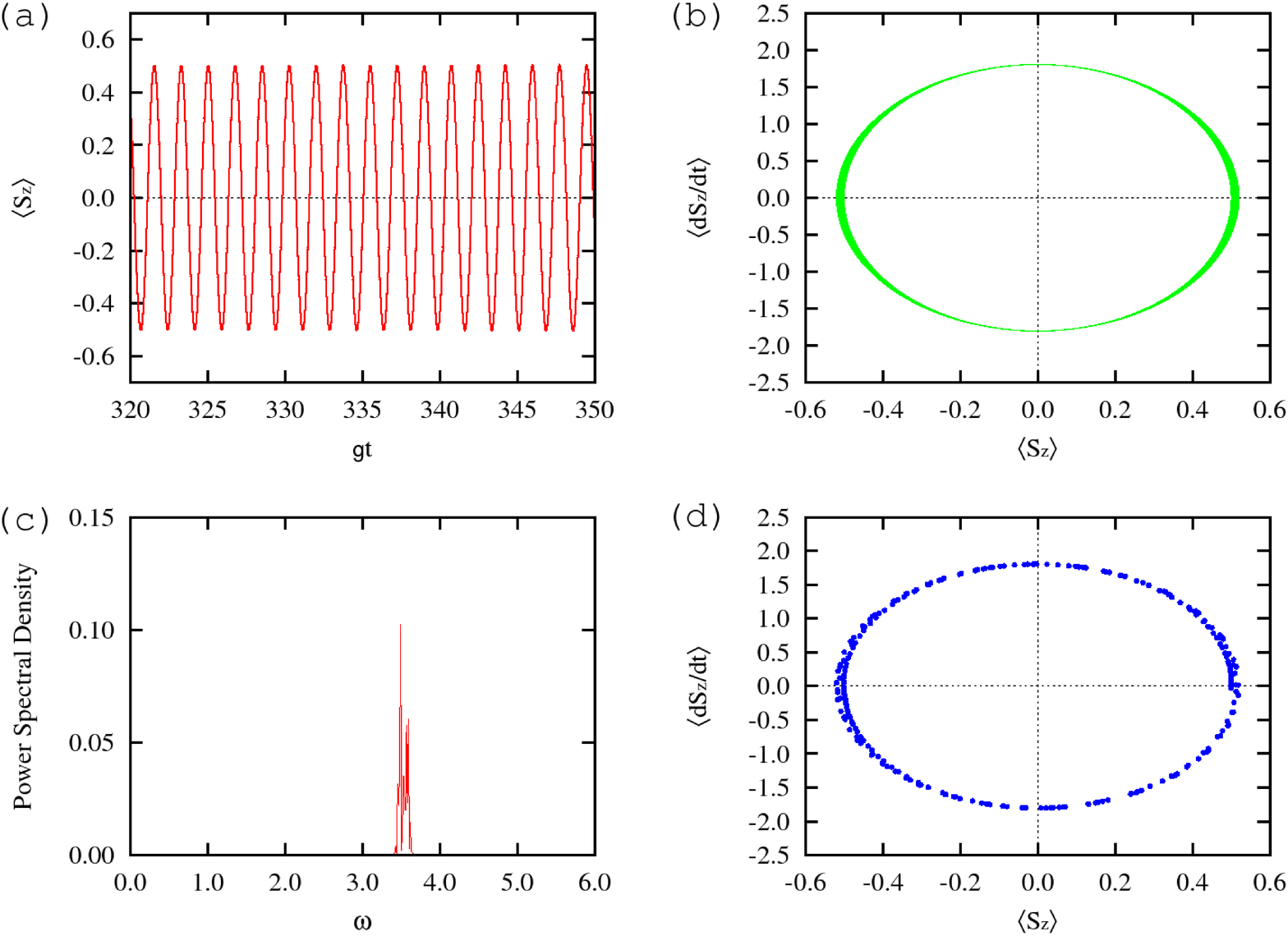}
\caption{\label{Fig04}Rabi frequency overtones modulating the cavity field.
Quasiperiodicity is observed. The parameters of the system are: $N=2.50$,
$g=1.00$, $\delta_0=1.00$, $\omega=0.01\,\Omega$, and $\avg{S_{z}(0)}=+1/2$. The
plots are: (a) time series, (b) phase plane, (c) power spectral density, and (d)
Poincar\'{e} section.}
\end{figure}

In Fig.~\ref{Fig04}, we observed another interesting behavior in the dynamics of
the system. This time $\omega=(1/m)\,\Omega$ with $m=100$. The ratio of $\omega$
to $\Omega $ is a fraction (with a positive integral denominator). Now the
modulation frequency and the Rabi frequency are no longer synchronized with each
other. Figure~\ref{Fig04}(a), displaying the time series for $\avg{S_{z}(t)}$,
apparently shows a single frequency of oscillation. However, the phase plot in
Fig.~\ref{Fig04}(b) reveals a small bandwidth around the single frequency which
is further confirmed in the PSD in Fig.~\ref{Fig04}(c), where there is no single
frequency showing up in the spectrum. The PS for this condition is displayed in
Fig.~\ref{Fig04}(d). This is typical of \textit{quasiperiodic} motion. The phase
trajectory goes around the phase plane over and over, never exactly repeating
itself. As the time increases, more points will show up in the PS, but never
coinciding with each other. Analyzing~\eqref{Eq:IHamiltonian} for $\delta =
\delta_0 \cos(\omega t)$, we find that the two levels in the system will cross
if $\cos(\omega t)=0$ or $t={n\pi}/{2\omega}$ ($n=1,3,5,\ldots$).

\begin{figure}[t]
\centering
\includegraphics[scale=0.16]{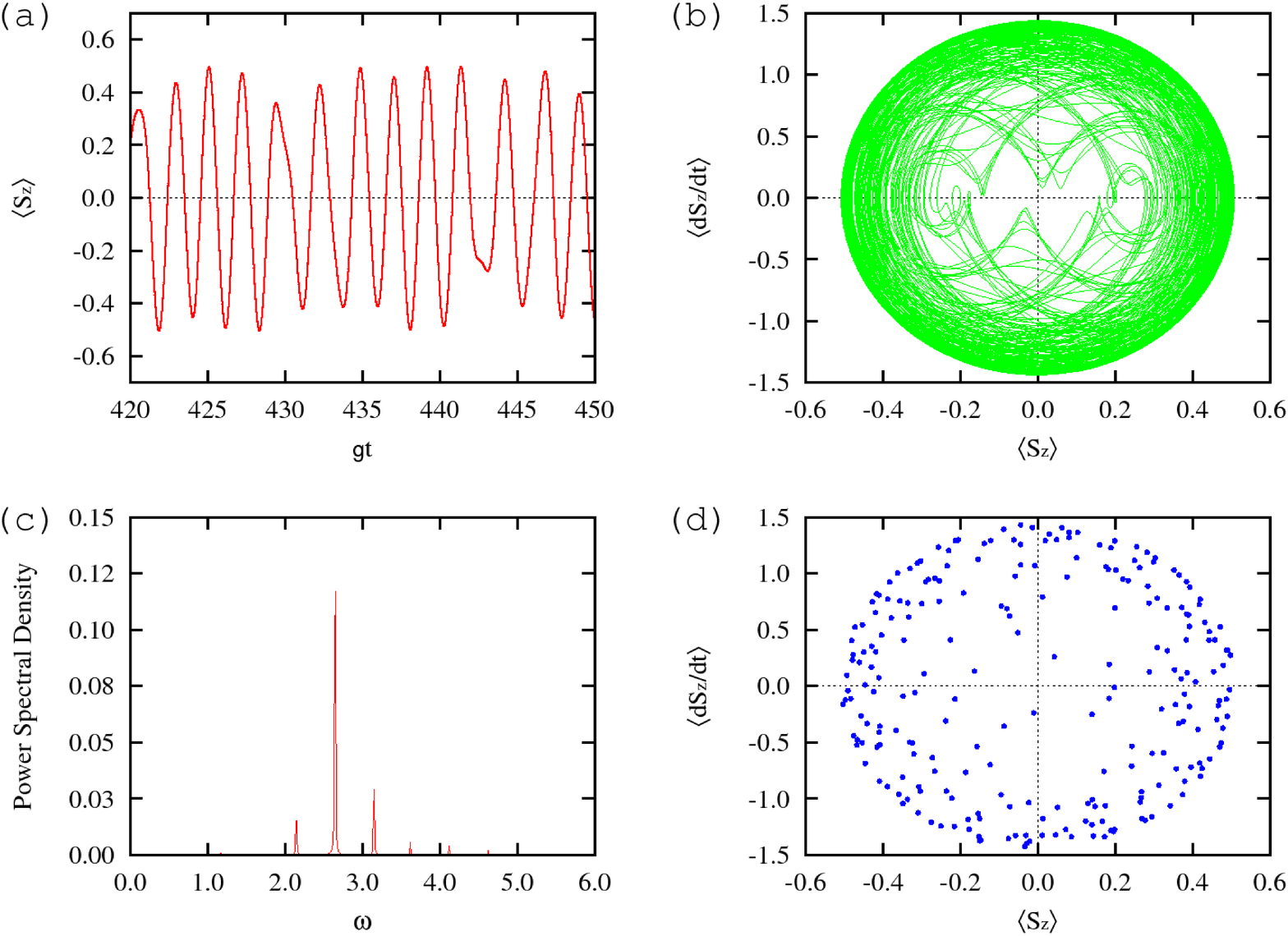}
\caption{\label{Fig05}Bichromatic modulation of the cavity field. The time
series plot shows an irregular pattern of modulation. Different frequencies
determine the dynamics of the system as shown in the power spectrum plot. The
parameters of the system are: $N=1.50$, $g=1.00$, $\delta_0=1.00$,
$\omega_1=\sqrt{7}$, $\omega_2=\sqrt{17}$, and $\avg{S_{z}(0)}=+1/2$. The plots
are: (a) time series, (b) phase plane, (c) power spectral density, and (d)
Poincar\'{e} section.}
\end{figure}

Finally, we considered bichromatic modulation of the cavity field by introducing
the expression $$\delta = \delta_0[\cos(\omega_1 t) + \cos(\omega_2t)].$$ The
parameters used were $N=1.50$, $g=1.00$, $\delta_0=1.00$, $\omega_1=\sqrt{7}$,
and $\omega_2=\sqrt{17}$, with a Rabi frequency $\Omega=3.00$. The atom was
initially in its excited state. In Fig.~\ref{Fig05}(a), the time series of
$\avg{S_{z}(t)}$ shows what can be considered a sinusoidal irregular modulation.
In Fig.~\ref{Fig05}(b), the phase plot displays \textit{quasiperiodic} behavior.
This is because the modulation frequencies selected are the square root of two
prime numbers, 7 and 17, and the ratio of the two frequencies
$\omega_2/\omega_1$, called the \textit{winding number}~\cite{Hand}, is
irrational. Hence their combination produces incommensurate dynamical evolution
of the physical variables $\avg{S_z}$ and $\avg{\dot{S}_z}$. The PSD
corresponding to the time series is shown in Fig.~\ref{Fig05}(c). There are
several frequencies in the power spectrum exhibiting the quasiperiodic behavior
of the system. The PS plot (Fig.~\ref{Fig05}(d)) shows how the motion of the
system never exactly repeat itself, filling the phase plane's area as time
progresses.

\begin{figure}[t]
\centering
\includegraphics[scale=0.16]{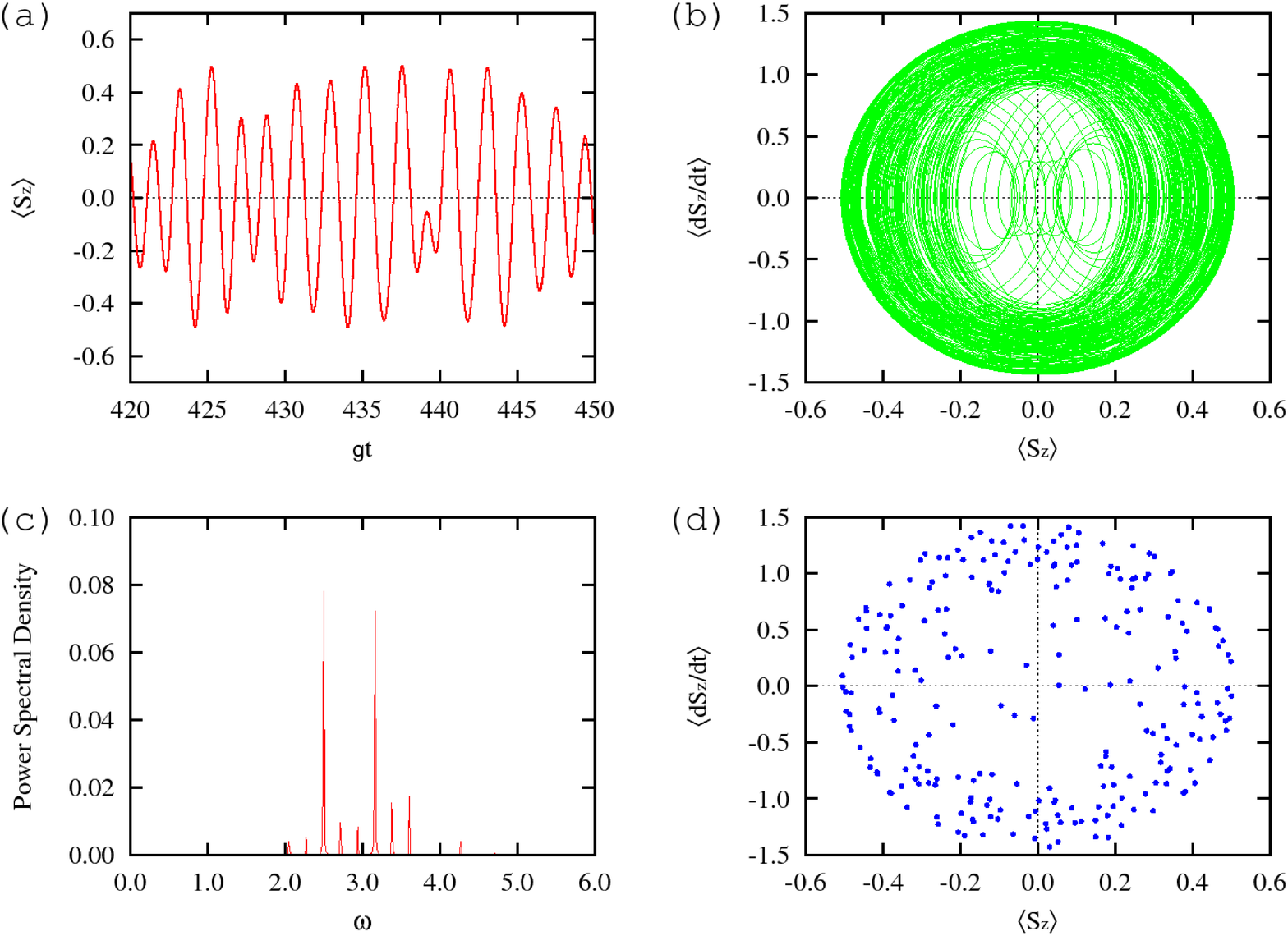}
\caption{\label{Fig06}Same as in Fig.~\ref{Fig05} but now the parameters of the
system are: $N=1.50$, $g=1.00$, $\delta_0=1.00$, $\omega_1=\sqrt{10}$,
$\omega_2=\sqrt{13}$, and $\avg{S_{z}(0)}=+1/2$. The plots are: (a) time series,
(b) phase plane, (c) power spectral density, and (d) Poincar\'{e} section.}
\end{figure}

Another interesting case of bichromatic modulation is presented in
Fig.~\ref{Fig06}. For this case, the parameters used were $N=1.50$, $g=1.00$,
$\delta_0=1.00$, $\omega_1=\sqrt{10}$, and $\omega_2=\sqrt{13}$, with a Rabi
frequency $\Omega=3.00$. The time series is shown in Fig.~\ref{Fig06}(a) and
exhibits a similar irregular modulation pattern to that observed in
Fig.~\ref{Fig05}(a). Again, this is because the ratio of the modulation
frequencies is an irrational number and incommensurate dynamical evolution
emerges. The phase plot in this situation brings out the quasiperiodic behavior
of the system. The phase trajectory clearly outlines two distinct regions; one
is an area densely populated by the states of the system, whereas the other is
rarely visited by the system (see Fig.~\ref{Fig06}(b)). The PSD in
Fig.~\ref{Fig06}(c) confirms this quasiperiodicity, where we observe a number of
different frequencies scattered in a relatively small portion of the spectrum.
Figure~\ref{Fig06}(d) shows the PS of the dynamical system. We observe how the
PS points cover the area of the phase plane as time moves forward.

We also studied the behavior of the system when the two modulated frequencies
$\omega_1$ and $\omega_2$ were overtones of the Rabi frequency $\Omega$ of the
system. As it happened before with the monochromatic case, only one
\textit{dominant} frequency emerged in the PSD plot for most of the cases. Two
or three marginal peaks eventually appeared in the plots, but they were
negligible compared with the single dominant frequency peak. We noted a slight
modulation in the time series for these cases (perhaps caused by those marginal
frequencies), but still a relative single frequency could be observed.
Unfortunately, the numerical algorithms were not stable and reliable enough for
the chosen parameters, so we decided not to proceed further in the study of this
particular case.

%% -----------------------------------------------------------------------------
%%                               CONCLUSIONS
%% -----------------------------------------------------------------------------
\section{\label{sec:summary}Summary}

In summary, we presented an alternative method used to study the interaction of
a two-level atom with a modulated quantized field of a cavity. This method
comprises different tools used to investigate nonlinear dynamical systems, like
the time series, phase plane, power spectrum, and Poincar\'{e} section plots.
With these instruments, we studied the dynamical evolution of a quantum system.

We described the interaction of the two-level atom with the cavity field using a
Jaynes-Cumming Hamiltonian. The derived Heisenberg equations of motion of the
system formed a set of three nonlinear first order differential equations.
Because of the \textit{stiffness} and complexity of the equations, we used
implicit Runge-Kutta, and semi-implicit Rosenbrock numerical methods to compute
the solutions. A great deal of educational value is obtained also from this part
of the work, since a simple algorithm for solving ODEs cannot be used here,
without compromising the accuracy and reliability of the solutions.

We explored and discussed three different scenarios: no modulation,
monochromatic, and bichromatic modulation of the cavity field. For the different
parameters that we used, we observed periodic and quasiperiodic behavior in the
population of the atomic states. We looked specially at those cases in which the
frequencies involved were integer, rational, and irrational numbers, as well as
overtones of the Rabi frequency of the system. When the ratio of the frequencies
was an irrational number, we observed quasiperiodic behavior of the system. For
frequencies that were overtones of the Rabi frequency of the system, we noted
just a single dominating frequency for the system.

The importance and final goal of this work are to help undergraduate students to
understand the dynamics of a two-level atom interacting with a quantized mode of
an optical cavity, when the cavity field is modulated. These goals are achieved
by working with tools used to explore nonlinear dynamical systems, and solving
numerically a set of equations of motion with specialized numerical algorithms.

%% -----------------------------------------------------------------------------
%%                             ACKNOWLEDGMENTS
%% -----------------------------------------------------------------------------
\begin{acknowledgments}
One of the authors (J.D.S.) gratefully acknowledge the School of Mathematical
and Natural Sciences at the University of Arkansas-Monticello (Grant
\#11-2225-5-M00) for providing funding and support for this work.
\end{acknowledgments}

%% -----------------------------------------------------------------------------
%%                                REFERENCES
%% -----------------------------------------------------------------------------
%\bibliography{cQED.bib}
%

\end{document}